\documentclass[superscriptaddress,aps,prl,twocolumn,showpacs,nofootinbib,longbibliography]{revtex4-1}
\usepackage{amsmath,amssymb,amsthm}
\usepackage{easybmat}
\usepackage[colorlinks=true,citecolor=blue,urlcolor=blue]{hyperref}
\usepackage[pdftex]{graphicx}
\usepackage{times,txfonts}
\usepackage{braket}
\usepackage{color}
\usepackage{natbib}

\newcommand{\be}{\begin{equation}}
\newcommand{\ee}{\end{equation}}
\newcommand{\ba}{\begin{eqnarray}}
\newcommand{\ea}{\end{eqnarray}}

\newcommand{\tr}{\operatorname{Tr}}

\begin{document}
\title{Revealing quantitative relation between simultaneous correlations in complementary bases and quantum steering for two qubit Bell diagonal states}   
 \author{C.  Jebarathinam} \email{jebarathinam@bose.res.in}
\affiliation{S. N. Bose National Centre for Basic Sciences, Salt Lake,
  Kolkata      700     106,      India}     \author{Aiman Khan}
\author{Som Kanjilal} \author{Dipankar Home} \affiliation{Center for Astroparticle
  Physics and Space Science (CAPSS),  Bose Institute, Block EN, Sector
  V,  Salt Lake,  Kolkata 700  091, India}

\date{\today}
\begin{abstract}
The present work is motivated by the question as to what aspect of correlation entailed by the two-qubit
state serves as the appropriate quantitative resource for  steering. To this end, considering Bell-diagonal states, suitable measures of simultaneous correlations in two and three complementary (mutually unbiased) bases are  identified as the relevant resources for quantum steering. Quantitative relations between appropriate measures of quantum steering and the corresponding measures of simultaneous correlations in complementary bases are demonstrated which ensure that for two qubit steerable Bell-diagonal states, higher value of simultaneous correlations in mutually unbiased bases necessarily implies higher degree of quantum steering, both for two and three setting steering scenarios.  
\end{abstract}


\maketitle

\textit{Introduction:-} The idea of quantum steering, one of the fundamental nonclassical features of quantum theory, was first suggested by Schrodinger in his response paper \cite{Sch35} to the  Einstein-Podolsky-Rosen argument \cite{EPR35}. This idea was formalized by Wiseman et al \cite{WJD07} stimulating extensive studies on quantum steering \cite{JWD07,SJW+10,BCW+12,Pus13,SNC14,BVQ+14,QVB14,UMG14,PW15,CBW+15,GA15,ZHC16}. Besides its foundational implications, usefulness of steerable states has been demonstrated in quantum information processing  applications such as one-sided device independent  quantum cryptography \cite{BCW+12}, randomness generation \cite{LTB+14} and sub-channel discrimination \cite{PW15}. These works have given  rise to the need for identifying an appropriate quantitative measure of steering. For this purpose, several measures of quantum steering have been proposed such as steerable weight \cite{SNC14}, steering robustness \cite{PW15}, steering fraction \cite{HLL16}, intrinsic steerability \cite{KWW17}, relative entropy of steering \cite{GA15,KW17} and steering cost \cite{DDJ+18}. Further, various works \cite{Pus13,SH18,QVC+15} have established that while entanglement is necessary for quantum steering, an entangled state in itself is not sufficient to ensure steering. This naturally gives rise to the question as to what particular aspect of quantum correlation, apart from entanglement, plays a crucial role in quantum steering. 

In this work, it is the above mentioned question that is addressed for the case of two-qubit Bell diagonal states by invoking the following two key ingredients:
\begin{itemize}
    \item As a quantitative measure of steering, we employ the steering measure proposed by Costa and Angelo (abbreviated as CA)  \cite{CA16} is employed which we choose because it is the only steering measure that has been shown to be amenable to a closed analytical expression for all two-qubit states.
    
    \item In order to bring out the feature of quantum correlations that is sought to be quantitatively related to the above mentioned steering measure, we take recourse to the recently proposed measures of simultaneous correlations in mutually unbiased bases (SCUMB) suggested by Guo and Wu \cite{GW14}.The basic idea underpinning such a measure is briefly explained below. 
\end{itemize}
  
  Considering a given measurement basis at Alice's end, correlations with respect to this basis can be characterized by the Holevo quantity \cite{NC00}, which is the upper bound to the maximum accessible information Bob can obtain about outcomes of Alice's measurement. We then consider the Holevo quantities corresponding to the bases which are mutually incompatible to the given basis and take the minimum of all such Holevo quantities - this minimum is regarded as the maximum amount of simultaneous correlation that is present in the corresponding set of mutually unbiased bases \cite{,WMC+14,GW14}. Thereafter, by varying the choice of the measurement basis, one can span the entire set of mutually unbiased bases and take the overall maximum of all such quantities mentioned earlier. This maximum is then interpreted as the maximum amount of simultaneous correlation that can be present in a given set of MUBs corresponding to any arbitrary choice of the measurement basis at Alice's side \cite{GW14}. Its non-vanishing value is thus taken to signify the persistence of correlations in sets of different incompatible bases as a fundamental quantum feature, and is regarded as the measure of simultaneous correlations in mutually unbiased bases (SCMUB) denoting a fundamental quantum feature. Importantly, this measure has been shown to be nonzero if and only if the state in question is a non-product state. Analytical expressions for this measure have also been obtained for the Bell-diagonal family of states \cite{WMC+14,GW14}. 

In this paper, using the steering measure proposed by CA \cite{CA16}, first for the two-setting steering scenario, it is demonstrated that there exists an analytical relationship between SCMUB and the CA steering measure derived using the two-setting linear steering inequality. We then show that SCMUB bears an analytical relationship with the corresponding steering measure in the three-setting scenario, too. These analytical relationships are such that for all steerable Bell-diagonal states corresponding to non-vanishing values of either two or three-setting steering measure, a higher value of SCMUB necessarily implies a higher degree of quantum steering.

This letter is organized as follows. First, we briefly discuss the core ideas of quantum steering and explain in detail the measures of steering proposed by CA. This is followed by discussing the relevant specifics of SCMUB and we demonstrate that both the two and three setting CA steering measures are, in fact, monotonically increasing functions of SCMUBs for any steerable Bell diagonal state, implying the significance of SCMUB as resource in the steering scenario. Finally, we conclude by indicating a few novel directions of study that  our work may open up.

\textit{Quantum steering:-} In a typical steering scenario, one of the two spatially separated parties, say Alice,  performs a set of black-box measurements (without any assumption about the measured state and the measurement device) that result in conditional states being prepared for the other party Bob, who, in turn, can perform state tomography to determine these conditional states. Bob will be convinced that Alice has indeed steered the preparation of the set of conditional states on his side if and only if the correlations between the Alice's measurement results and the set of conditional states on Bob's side \textit{cannot} be explained by a local hidden state (LHS) model \cite{WJD07}. If Alice and Bob share a separable state, it has been shown that the above mentioned correlations can always be explained using a LHS model \cite{PW15}. Thus, separable states cannot be useful for steering, whence demonstration of steering necessarily requires the presence of entanglement between the parties.

Steering can also be viewed operationally within the framework of quantum information theory as the certification of entanglement in a one-sided device-independent manner \cite{WJD07,QVC+15}, as the steering scenario corresponds to one party performing untrusted measurements while the other party performs trusted measurements. On the other hand, certification of entanglement through demonstration of Bell nonlocality \cite{Bel64} is fully-device-independent \cite{BCP+14}. Also, to be noted that steering inequalities have been formulated \cite{CJW+09,ZHC16}, analogous to Bell inequalities, which can certify the presence of steering. In a nutshell, steering lies intermediate between entanglement and Bell nonlocality: quantum states that exhibit Bell nonlocality form a subset of EPR steerable states which, in turn, form a subset of entangled states \cite{WJD07,QVC+15}.

\textit{Measure of steering proposed by Costa and Angelo (CA):-} A general $n$-setting steering scenario comprises Alice performing $n$-dichotomic black-box measurements $A_k \in \{-1,1\}$, where $k=1,2,3, \cdots n$, and Bob performing $n$-two outcome projective measurements 
$B_k=\hat{b}_k \cdot \vec{\sigma}$ along the unit directions $\hat{b}_k$, here $\vec{\sigma}$  is the vector of Pauli matrices. For such $n$-setting scenario, the linear steering inequality formulated by Cavalcanti, Jones, Wiseman, and Reid (CJWR) \cite{CJW+09} as the sum of $k$ bilinear expectation values is given by \cite{CJW+09}:
\be
\label{FJWR}
F_n(\rho,\mu):=\frac{1}{\sqrt{n}} \left|\sum^n_{k=1} \braket{A_k\otimes B_k}\right| \le 1, 
\ee
where $\rho$ is the given bipartite quantum state,  $\mu=\{A_k;B_k\}^n_{k=1}$ refers to the measurement settings of Alice and Bob and $\braket{A_k\otimes B_k}=\tr(\rho(A_k\otimes B_k))$. Violation of any of these inequalities certifies steering using a given quantum state.  

Here our attention is confined to the two-qubit case. CA \cite{CA16} defined the following class of steerability measures that seek to quantify steerability of two-qubit states  via the maximal violation of the CJWR linear steering  inequalities (\ref{FJWR}):
\be
S_n(\rho)=\max\left\{0,\frac{F_n(\rho)-1}{F^{max}_n-1}\right\}, \label{SQ}
\ee
where 
\be
F_n(\rho)=\max_{\mu} F_n(\rho,\mu),
\ee
and 
$F^{max}_n=\max_{\rho}F_n(\rho)$. 
It is readily seen that any $S_n$ is non-zero if and only if there is a corresponding violation of the linear steering inequality, that is, $F_n >1$.

  For the two- and three-setting scenarios, CA further derived, for any two-qubit state, closed analytical expressions for the steering measures $S_2$ and $S_3$ which are functions of components of the correlation vector 
$\vec{c}:=\{c_1,c_2,c_3\}$ of the density matrix describing the quantum state, with the two-qubit state parametrized in the following manner:
\be
\zeta=\frac{1}{4}\left(\openone \otimes \openone + \vec{a} \cdot \vec{\sigma} \otimes \openone 
+ \openone \otimes \vec{b} \cdot \vec{\sigma} +\sum^3_{i=1} c_i \sigma_i \otimes \sigma_i \right), \label{can02q}
\ee
where $\openone$ is the $2 \times 2$ identity matrix and $\{\vec{a},\vec{b},\vec{c}\} \in \mathbf{R}^3$ are vectors with norm less than or equal to unity and $\vec{a}^2+\vec{b}^2+\vec{c}^2\le3$. In Ref.  \cite{Luo08}  it was shown that any two-qubit state, up to local unitary transformations, can be reduced to the above form, making the above parametrization applicable to all two-qubit states.

Specifically, $S_2$ and $S_3$ have been evaluated to be the following \cite{CA16}:
\be \label{S2m}
S_2(\zeta)=\mbox{max}\left\{0,\frac{F_2(\zeta)-1}{F^{max}_2-1}\right\},\,\,\, F_2(\zeta) = \sqrt{c^2-c^2_{\min}}  
\ee
and 
\be \label{S3m}
S_3(\zeta)=\mbox{max}\left\{0,\frac{F_3(\zeta)-1}{F^{max}_3-1}\right\} ,\,\,\, F_3(\zeta) = c  
\ee
respectively, where $c=\sqrt{\vec{c}.\vec{c}}$ is the norm of correlation vector $\vec{c}$ and $c_{\min}=\min\{|c_1|,|c_2|,|c_3|\}$. It is these two expressions which will be used later for relating steering measures with SCMUB.

\textit{Measures of simultaneous correlations in mutually unbiased bases:-} Let Alice and Bob share a bipartite state $\rho_{AB}$. Dimension of local Hilbert spaces corresponding to Alice and Bob are taken to be $d$. Let Alice encode the information about a classical random variable which takes $d$ number of values using local measurement in the basis $\{\ket{a_{i}}_{A}|i \in (1,2,...,d)\}$. Upper bound on the accessible information about the encoded classical random variable available to Bob is given by
\begin{align}
    \label{s4m}
    \mathcal{C}_{1} & = \max_{\{\Pi_{i}^{A}\}}S\big(\sum_{i}p_{i}\rho^{B}_{i}\big)- \sum_{i}p_{i}S\big(\rho^{B}_{i}\big)\\
    \label{s5m}
    & = \max_{\{\Pi_{i}^{A}\}}\chi(\rho_{AB},\{\Pi_{i}^{A}\})
\end{align}
where $S(\rho)=-Tr[\rho\log\rho]$ is the von Neumann entropy,  $\rho^{B}_{i}=_{A}\langle a_{i}|\rho_{AB}| a_{i}\rangle_{A}/p_{i}$, $p_{i}=Tr[_{A}\langle a_{i}|\rho_{AB}| a_{i}\rangle_{A}]$, $\Pi_{i}^{A}=\ket{a_{i}}_{A}\bra{a_{i}}$ and $\chi(\rho_{AB},\{\Pi_{i}^{A}\})$ is called the Holevo quantity pertaining to the ensemble $\{p_{i},\rho^{B}_{i}\}$ corresponding to Bob's end \cite{HV01}. Note that $\mathcal{C}_{1}$ as given by Eqs. (\ref{s4m}) and (\ref{s5m}) is the maximum accessible information about a classical random variable contained in an arbitrary measurement basis $\{\ket{a_{i}}_{A}|i \in (1,2,...,d)\}$ when $\rho_{AB}$ is used as a resource channel. Therefore, one can interpret $\mathcal{C}_{1}$ as the maximum classical information contained in a state $\rho_{AB}$ \cite{WMC+14}.

A fundamental feature of quantum mechanics is the existence of mutually unbiased bases. Two sets of complete bases, say $\{\ket{a_{i}^{1}}\}$ and $\{\ket{a_{j}^{2}}\}$ 
in Hilbert space of dimension $d$ are defined to be mutually unbiased if and only if $$ |\langle a_{i}^{1}|a_{j}^{2}\rangle|=\frac{1}{\sqrt{d}}.$$ Physically, this can be interpreted in the following way: Suppose the observer is provided a system in one of the basis kets, say $\ket{a_{m}^{1}}$, and measurement is performed given by projectors $\{\Pi_n=\ket{a_n^2}\bra{a_n^2}\}$. Then the resulting state would be projected with equal probability onto each of the basis kets of the other basis $\{\ket{a_n^2}\}$.  

Here we note that quantum phenomenon such as violation of the Bell inequality is related to the existence of simultaneous correlations in observed properties in mutually unbiased bases. It is this idea that has been used to propose a series of measures \cite{WMC+14,GW14} that seek to capture the quantumness of correlations through the persistence of correlations in mutually unbiased bases corresponding to a bipartite state $\rho_{AB}$.  To illustrate this, we denote by $\Omega_{A}$ the set of all pairs of bases that are mutually unbiased with each other, that is to say:

\begin{align}
    \Omega_{A} & := \{ \{ \{\ket{a_i^1}_{A}\}, \{\ket{a_j^2}_{A}\} \}: |_{A}\braket{a_i^1|a_j^2}_{A}| = \frac{1}{\sqrt{d}} \nonumber\\
    & \forall i,j \in (1,2,...,d) \}.
\end{align}

One can now define the quantity $\mathcal{C}_2$ as the maximum amount of simultaneous correlations that exist in any given pair of mutually unbiased bases (SCMUB), that is \cite{GW14}:

\begin{equation} \label{C2def}
    \mathcal{C}_2 = \max_{\Pi_1^A,\Pi_2^A\in\Omega} \min [\chi(\rho_{AB},\{\Pi_1^A\}),\chi(\rho_{AB},\{\Pi_2^A)\}],
\end{equation}
where $\{\Pi_i^A\}$ represents the basis of measurement in Alice's local Hilbert space. This definition can be easily generalized by taking more than two bases at a time. For a bipartite quantum state $\rho_{AB}$ with the local Hilbert space dimension in Alice's side being $d$, one can define the quantity as in Eq. (\ref{C2def}) with $m$ mutually unbiased bases, here $3 \le m \le d+1$. For $d=2$, the measure of SCMUB defined as in Eq. (\ref{C2def}) cannot be generalized with more than three bases since for qubit systems there cannot be more than three mutually unbiased bases. Next, similar to the quantity $\mathcal{C}_2$, one can define $\mathcal{C}_3$ as follows  \cite{GW14}:

\begin{equation} \label{C3def}
    \mathcal{C}_3 = \max_{\Pi_1^A,\Pi_2^A,\Pi_3^A\in\Lambda_{A}} \min [\chi(\rho_{AB},\{\Pi_1^A\}),\chi(\rho_{AB},\{\Pi_2^A\}),\chi(\rho_{AB},\{\Pi_3^A\})],
\end{equation}
where the set of all triads of mutually unbiased bases in Alice's Hilbert space is denoted by $\Lambda_{A}$ as
\begin{align}
    \Lambda_{A} & := \{ \{ \{\ket{a_i^1}_{A}\}, \{\ket{a_j^2}_{A}, \{\ket{a_k^3}_{A}\} \}: |_{A}\braket{a_i^1|a_j^2}_{A}| = |_{A}\braket{a_j^2|a_k^3}_{A}|\nonumber\\
    & =|_{A}\braket{a_k^3|a_i^1}_{A}| = \frac{1}{\sqrt{d}} \hspace{5mm} \forall i,j,k \in (1,2,...d)\}.
\end{align}

Thus, having defined appropriate measures of simultaneous correlation in mutually unbiased bases (SCMUB), we will now derive the central result of this paper showing explicit analytical relationship between steering measures and SCMUB in the two- and three-setting steering scenarios.

\textit{Relating steering measures and measures of complementary correlations:-}
For the two-qubit Bell-diagonal family of states, the measure in two mutually unbiased bases of simultaneous correlations $\mathcal{C}_2$ has been evaluated to be the following \cite{GW14}:
\be \label{C2BD}
\mathcal{C}_2(\tau)=1-h\left(\frac{1+\sqrt{(c^2-c^2_{\min})/2}}{2}\right),
\ee
with $h(x)\equiv -x \log_2x - (1-x) \log_2(1-x)$.
Using Eqs. (\ref{S2m}) and (\ref{C2BD}), it is seen that $C_2(\tau)$ and $F_2(\tau)$ are related to each other as follows:
\be
\label{C2F2}
\mathcal{C}_2(\tau)=1-h\left(\frac{1+F_2(\tau)/\sqrt{2}}{2}\right).
\ee
Now, note that, if a function $f(x)$ continuous and differentiable in any given interval $a \leq x \leq b$, then $f(x)$ is monotonically increasing function of $x$ provided the derivative of $f(x)$ with respect to $x$ is not negative, i.e., $\frac{df(x)}{dx} \geq 0$ for all $x \in [a,b]$. By differentiating the left hand side of Eq. (\ref{C2F2}) with respect to $F_{2}(\tau)$, it  can be checked that $\frac{d\mathcal{C}_{2}}{dF_{2}} \geq 0$ for all values of $F_{2}$ which implies that $\mathcal{C}_{2}$ is a monotonically increasing function of $F_{2}$.

Further, note that, it follows from the definition of steering measure $S_n$ (Eq. (\ref{SQ})) that as long as there is a violation of the steering inequality in the two-setting scenario, i.e. $F_2>1$, $S_2$ is a monotonically increasing function of $F_2$. Then from the monotonic relationship between $\mathcal{C}_{2}$ and $F_{2}$ given by Eq. (\ref{C2F2}), one can also infer the monotonicity of $\mathcal{C}_{2}$ with respect to $S_{2}$ for the two-qubit Bell-diagonal states. 

Here it should also be noted that, from Eq. (\ref{SQ}) one can infer that when $S_{2}$ vanishes $F_{2}$ becomes zero. Now, if we insert $F_{2}=0$ in Eq. (\ref{C2F2}), it follows that $\mathcal{C}_{2}=0$. Therefore, when $\mathcal{C}_{2}$ vanishes, $S_{2}$ vanishes, too. 

Similarly, for the three mutually unbiased bases, the measure of simultaneous correlations $\mathcal{C}_3$ (defined by Eq. (\ref{C3def})) for the Bell-diagonal states has been found to be \cite{GW14}:
\be \label{C3BD}
C_3(\tau)=1-h\left(\frac{1+c/\sqrt{3}}{2}\right).
\ee
Then the following relationship can be easily obtained using Eqs. (\ref{S3m}) and (\ref{C3BD}):
\be \label{C3BDF3}
C_3(\tau)=1-h\left(\frac{1+F_3(\tau)/\sqrt{3}}{2}\right).
\ee
 Similar to the earlier two-setting case, it  can be checked from Eq. (\ref{C3BDF3}) that the first derivative of $C_3$ with respect to $F_3$ is non-negative, i.e., $\frac{d\mathcal{C}_{3}}{dF_{3}}\geq 0$ for all values of $F_{3}$;  i.e., in the three-setting case too, $C_3$ is a monotonically increasing function of $F_{3}$. Then the above Eq. (\ref{C3BDF3}), in conjunction with the definition of $S_n$ in Eq. (\ref{SQ}), implies that for the two-qubit Bell-diagonal states, the measure of steering $S_3(\tau)$ is a monotonically increasing function of $C_3(\tau)$ in the region $S_3>0$ (i.e., $F_3>1$ that corresponds to violation of the three-setting steering inequality). Here also, if $\mathcal{C}_{3}$ vanishes, then $S_{3}$ vanishes.

\textit{Concluding remarks:-}
In conclusion, let us summarize the key features of this work. Appropriate quantitative resource for quantum steering with respect to the steering quantifier proposed in Ref. \cite{CA16} has been identified by demonstrating that for the Bell-diagonal two-qubit states, such steering quantifiers $F_{2}$ and $F_{3}$ are analytically and monotonically related to the degree of simultaneous correlations in mutually unbiased bases as quantified by $C_2$ and $C_3$ in the two and three-setting scenarios respectively. Such quantitative relationships reinforce the link between measurement incompatibility and steering that has been indicated through a one-to-one correspondence between the mutual incompatibility of Alice's measurements and whether Bob's state can be steered by Alice \cite{QVB14,UMG14}. 
The importance of such a connection between steering and measurement incompatibility has also been brought out through the construction of measure of steering based on measurement incompatibility \cite{UBG+15} which provides a lower bound to the corresponding measure of measurement incompatibility \cite{CBL+16}. 


The identification made in this paper of what aspect of correlations embodied in the bipartite quantum state quantifies quantum steering can thus play a key role in facilitating exploration of its foundational implications as well as its applications in quantum information processing, the potentiality of which can be studied through appropriate examples such as the sub-channel discrimination problem which refers to discrimination of the branches in a quantum evolution; interestingly, it has been shown that what is defined as steering robustness quantifies quantum advantage for the task of subchannel discrimination using steerable states \cite{PW15}. In light of this result and our present work, it should therefore be interesting to study the relationship between steering robustness and SCUMB. Similarly, the link between steering and one-sided device independent quantum key distribution, as pointed out by Branciard et al \cite{BCW+12}, together with the results of our paper may enable identification of the appropriate resource in such quantum cryptographic protocols.

Note that our demonstration of the analytical relationships between steering measures and SCMUB  measures has been essentially for the two-qubit Bell-diagonal states, and a natural line of further inquiry would be to investigate whether a similar relationship would hold for a general two-qubit state as resource.

Finally, it needs to be noted that while the steering quantifiers $S_n$ with $n=2,3$ have closed analytic forms for all two-qubit states, it should be worthwhile to examine their validity in the wider context of the resource theory of steering \cite{GA15}. In particular, this would entail studying whether these measures are non-increasing with respect to the allowed operations (local operations assisted by one-way classical
communications from the trusted side to the black-box side) in the steering scenario. Such an investigation would help to bring out the foundational implications of steering measures $S_n$ as resource, and thereby through our work, that of SCMUB.

\textit{Acknowledgements:-} 
DH thanks Urbasi Sinha for drawing attention to Ref. \cite{GW14}  which helped to stimulate the present work. 
We also thank H. M. Wiseman for helpful and encouraging comments on the earlier version.
CJ acknowledges S. N. Bose Centre, Kolkata for the postdoctoral fellowship. SK acknowledges the support of a fellowship from Bose Institute, Kolkata.

\bibliography{JM}

\begin{thebibliography}{33}%
\makeatletter
\providecommand \@ifxundefined [1]{%
 \@ifx{#1\undefined}
}%
\providecommand \@ifnum [1]{%
 \ifnum #1\expandafter \@firstoftwo
 \else \expandafter \@secondoftwo
 \fi
}%
\providecommand \@ifx [1]{%
 \ifx #1\expandafter \@firstoftwo
 \else \expandafter \@secondoftwo
 \fi
}%
\providecommand \natexlab [1]{#1}%
\providecommand \enquote  [1]{``#1''}%
\providecommand \bibnamefont  [1]{#1}%
\providecommand \bibfnamefont [1]{#1}%
\providecommand \citenamefont [1]{#1}%
\providecommand \href@noop [0]{\@secondoftwo}%
\providecommand \href [0]{\begingroup \@sanitize@url \@href}%
\providecommand \@href[1]{\@@startlink{#1}\@@href}%
\providecommand \@@href[1]{\endgroup#1\@@endlink}%
\providecommand \@sanitize@url [0]{\catcode `\\12\catcode `\$12\catcode
  `\&12\catcode `\#12\catcode `\^12\catcode `\_12\catcode `\%12\relax}%
\providecommand \@@startlink[1]{}%
\providecommand \@@endlink[0]{}%
\providecommand \url  [0]{\begingroup\@sanitize@url \@url }%
\providecommand \@url [1]{\endgroup\@href {#1}{\urlprefix }}%
\providecommand \urlprefix  [0]{URL }%
\providecommand \Eprint [0]{\href }%
\providecommand \doibase [0]{http://dx.doi.org/}%
\providecommand \selectlanguage [0]{\@gobble}%
\providecommand \bibinfo  [0]{\@secondoftwo}%
\providecommand \bibfield  [0]{\@secondoftwo}%
\providecommand \translation [1]{[#1]}%
\providecommand \BibitemOpen [0]{}%
\providecommand \bibitemStop [0]{}%
\providecommand \bibitemNoStop [0]{.\EOS\space}%
\providecommand \EOS [0]{\spacefactor3000\relax}%
\providecommand \BibitemShut  [1]{\csname bibitem#1\endcsname}%
\let\auto@bib@innerbib\@empty
\bibitem [{\citenamefont {Schrodinger}(1935)}]{Sch35}%
  \BibitemOpen
  \bibfield  {author} {\bibinfo {author} {\bibfnamefont {E.}~\bibnamefont
  {Schrodinger}},\ }\bibfield  {title} {\enquote {\bibinfo {title} {Discussion
  of probability relations between separated systems},}\ }\href {\doibase
  10.1017/S0305004100013554} {\bibfield  {journal} {\bibinfo  {journal} {Math.
  Proc. Cambridge Philos. Soc.}\ }\textbf {\bibinfo {volume} {31}},\ \bibinfo
  {pages} {555--563} (\bibinfo {year} {1935})}\BibitemShut {NoStop}%
\bibitem [{\citenamefont {Einstein}\ \emph {et~al.}(1935)\citenamefont
  {Einstein}, \citenamefont {Podolsky},\ and\ \citenamefont {Rosen}}]{EPR35}%
  \BibitemOpen
  \bibfield  {author} {\bibinfo {author} {\bibfnamefont {A.}~\bibnamefont
  {Einstein}}, \bibinfo {author} {\bibfnamefont {B.}~\bibnamefont {Podolsky}},
  \ and\ \bibinfo {author} {\bibfnamefont {N.}~\bibnamefont {Rosen}},\
  }\bibfield  {title} {\enquote {\bibinfo {title} {Can quantum-mechanical
  description of physical reality be considered complete?}}\ }\href {\doibase
  10.1103/PhysRev.47.777} {\bibfield  {journal} {\bibinfo  {journal} {Phys.
  Rev.}\ }\textbf {\bibinfo {volume} {47}},\ \bibinfo {pages} {777--780}
  (\bibinfo {year} {1935})}\BibitemShut {NoStop}%
\bibitem [{\citenamefont {Wiseman}\ \emph {et~al.}(2007)\citenamefont
  {Wiseman}, \citenamefont {Jones},\ and\ \citenamefont {Doherty}}]{WJD07}%
  \BibitemOpen
  \bibfield  {author} {\bibinfo {author} {\bibfnamefont {H.~M.}\ \bibnamefont
  {Wiseman}}, \bibinfo {author} {\bibfnamefont {S.~J.}\ \bibnamefont {Jones}},
  \ and\ \bibinfo {author} {\bibfnamefont {A.~C.}\ \bibnamefont {Doherty}},\
  }\bibfield  {title} {\enquote {\bibinfo {title} {Steering, entanglement,
  nonlocality, and the einstein-podolsky-rosen paradox},}\ }\href {\doibase
  10.1103/PhysRevLett.98.140402} {\bibfield  {journal} {\bibinfo  {journal}
  {Phys. Rev. Lett.}\ }\textbf {\bibinfo {volume} {98}},\ \bibinfo {pages}
  {140402} (\bibinfo {year} {2007})}\BibitemShut {NoStop}%
\bibitem [{\citenamefont {Jones}\ \emph {et~al.}(2007)\citenamefont {Jones},
  \citenamefont {Wiseman},\ and\ \citenamefont {Doherty}}]{JWD07}%
  \BibitemOpen
  \bibfield  {author} {\bibinfo {author} {\bibfnamefont {S.~J.}\ \bibnamefont
  {Jones}}, \bibinfo {author} {\bibfnamefont {H.~M.}\ \bibnamefont {Wiseman}},
  \ and\ \bibinfo {author} {\bibfnamefont {A.~C.}\ \bibnamefont {Doherty}},\
  }\bibfield  {title} {\enquote {\bibinfo {title} {Entanglement,
  einstein-podolsky-rosen correlations, bell nonlocality, and steering},}\
  }\href {\doibase 10.1103/PhysRevA.76.052116} {\bibfield  {journal} {\bibinfo
  {journal} {Phys. Rev. A}\ }\textbf {\bibinfo {volume} {76}},\ \bibinfo
  {pages} {052116} (\bibinfo {year} {2007})}\BibitemShut {NoStop}%
\bibitem [{\citenamefont {Saunders~D.}(2010)}]{SJW+10}%
  \BibitemOpen
  \bibfield  {author} {\bibinfo {author} {\bibfnamefont {Wiseman H. Pryde~G.}\
  \bibnamefont {Saunders~D.}, \bibfnamefont {Jones~S.}},\ }\bibfield  {title}
  {\enquote {\bibinfo {title} {Experimental epr-steering using bell-local
  states},}\ }\href {\doibase 10.1038/nphys1766} {\bibfield  {journal}
  {\bibinfo  {journal} {Nat. Phys}\ }\textbf {\bibinfo {volume} {6}},\ \bibinfo
  {pages} {845} (\bibinfo {year} {2010})}\BibitemShut {NoStop}%
\bibitem [{\citenamefont {Branciard}\ \emph {et~al.}(2012)\citenamefont
  {Branciard}, \citenamefont {Cavalcanti}, \citenamefont {Walborn},
  \citenamefont {Scarani},\ and\ \citenamefont {Wiseman}}]{BCW+12}%
  \BibitemOpen
  \bibfield  {author} {\bibinfo {author} {\bibfnamefont {Cyril}\ \bibnamefont
  {Branciard}}, \bibinfo {author} {\bibfnamefont {Eric~G.}\ \bibnamefont
  {Cavalcanti}}, \bibinfo {author} {\bibfnamefont {Stephen~P.}\ \bibnamefont
  {Walborn}}, \bibinfo {author} {\bibfnamefont {Valerio}\ \bibnamefont
  {Scarani}}, \ and\ \bibinfo {author} {\bibfnamefont {Howard~M.}\ \bibnamefont
  {Wiseman}},\ }\bibfield  {title} {\enquote {\bibinfo {title} {One-sided
  device-independent quantum key distribution: Security, feasibility, and the
  connection with steering},}\ }\href {\doibase 10.1103/PhysRevA.85.010301}
  {\bibfield  {journal} {\bibinfo  {journal} {Phys. Rev. A}\ }\textbf {\bibinfo
  {volume} {85}},\ \bibinfo {pages} {010301 (R)} (\bibinfo {year}
  {2012})}\BibitemShut {NoStop}%
\bibitem [{\citenamefont {Pusey}(2013)}]{Pus13}%
  \BibitemOpen
  \bibfield  {author} {\bibinfo {author} {\bibfnamefont {Matthew~F.}\
  \bibnamefont {Pusey}},\ }\bibfield  {title} {\enquote {\bibinfo {title}
  {Negativity and steering: A stronger peres conjecture},}\ }\href {\doibase
  10.1103/PhysRevA.88.032313} {\bibfield  {journal} {\bibinfo  {journal} {Phys.
  Rev. A}\ }\textbf {\bibinfo {volume} {88}},\ \bibinfo {pages} {032313}
  (\bibinfo {year} {2013})}\BibitemShut {NoStop}%
\bibitem [{\citenamefont {Skrzypczyk}\ \emph {et~al.}(2014)\citenamefont
  {Skrzypczyk}, \citenamefont {Navascu\'es},\ and\ \citenamefont
  {Cavalcanti}}]{SNC14}%
  \BibitemOpen
  \bibfield  {author} {\bibinfo {author} {\bibfnamefont {Paul}\ \bibnamefont
  {Skrzypczyk}}, \bibinfo {author} {\bibfnamefont {Miguel}\ \bibnamefont
  {Navascu\'es}}, \ and\ \bibinfo {author} {\bibfnamefont {Daniel}\
  \bibnamefont {Cavalcanti}},\ }\bibfield  {title} {\enquote {\bibinfo {title}
  {Quantifying einstein-podolsky-rosen steering},}\ }\href {\doibase
  10.1103/PhysRevLett.112.180404} {\bibfield  {journal} {\bibinfo  {journal}
  {Phys. Rev. Lett.}\ }\textbf {\bibinfo {volume} {112}},\ \bibinfo {pages}
  {180404} (\bibinfo {year} {2014})}\BibitemShut {NoStop}%
\bibitem [{\citenamefont {Bowles}\ \emph {et~al.}(2014)\citenamefont {Bowles},
  \citenamefont {V\'ertesi}, \citenamefont {Quintino},\ and\ \citenamefont
  {Brunner}}]{BVQ+14}%
  \BibitemOpen
  \bibfield  {author} {\bibinfo {author} {\bibfnamefont {Joseph}\ \bibnamefont
  {Bowles}}, \bibinfo {author} {\bibfnamefont {Tam\'as}\ \bibnamefont
  {V\'ertesi}}, \bibinfo {author} {\bibfnamefont {Marco~T\'ulio}\ \bibnamefont
  {Quintino}}, \ and\ \bibinfo {author} {\bibfnamefont {Nicolas}\ \bibnamefont
  {Brunner}},\ }\bibfield  {title} {\enquote {\bibinfo {title} {One-way
  einstein-podolsky-rosen steering},}\ }\href {\doibase
  10.1103/PhysRevLett.112.200402} {\bibfield  {journal} {\bibinfo  {journal}
  {Phys. Rev. Lett.}\ }\textbf {\bibinfo {volume} {112}},\ \bibinfo {pages}
  {200402} (\bibinfo {year} {2014})}\BibitemShut {NoStop}%
\bibitem [{\citenamefont {Quintino}\ \emph {et~al.}(2014)\citenamefont
  {Quintino}, \citenamefont {V\'ertesi},\ and\ \citenamefont
  {Brunner}}]{QVB14}%
  \BibitemOpen
  \bibfield  {author} {\bibinfo {author} {\bibfnamefont {Marco~T\'ulio}\
  \bibnamefont {Quintino}}, \bibinfo {author} {\bibfnamefont {Tam\'as}\
  \bibnamefont {V\'ertesi}}, \ and\ \bibinfo {author} {\bibfnamefont {Nicolas}\
  \bibnamefont {Brunner}},\ }\bibfield  {title} {\enquote {\bibinfo {title}
  {Joint measurability, einstein-podolsky-rosen steering, and bell
  nonlocality},}\ }\href {\doibase 10.1103/PhysRevLett.113.160402} {\bibfield
  {journal} {\bibinfo  {journal} {Phys. Rev. Lett.}\ }\textbf {\bibinfo
  {volume} {113}},\ \bibinfo {pages} {160402} (\bibinfo {year}
  {2014})}\BibitemShut {NoStop}%
\bibitem [{\citenamefont {Uola}\ \emph {et~al.}(2014)\citenamefont {Uola},
  \citenamefont {Moroder},\ and\ \citenamefont {G\"uhne}}]{UMG14}%
  \BibitemOpen
  \bibfield  {author} {\bibinfo {author} {\bibfnamefont {Roope}\ \bibnamefont
  {Uola}}, \bibinfo {author} {\bibfnamefont {Tobias}\ \bibnamefont {Moroder}},
  \ and\ \bibinfo {author} {\bibfnamefont {Otfried}\ \bibnamefont {G\"uhne}},\
  }\bibfield  {title} {\enquote {\bibinfo {title} {Joint measurability of
  generalized measurements implies classicality},}\ }\href {\doibase
  10.1103/PhysRevLett.113.160403} {\bibfield  {journal} {\bibinfo  {journal}
  {Phys. Rev. Lett.}\ }\textbf {\bibinfo {volume} {113}},\ \bibinfo {pages}
  {160403} (\bibinfo {year} {2014})}\BibitemShut {NoStop}%
\bibitem [{\citenamefont {Piani}\ and\ \citenamefont {Watrous}(2015)}]{PW15}%
  \BibitemOpen
  \bibfield  {author} {\bibinfo {author} {\bibfnamefont {Marco}\ \bibnamefont
  {Piani}}\ and\ \bibinfo {author} {\bibfnamefont {John}\ \bibnamefont
  {Watrous}},\ }\bibfield  {title} {\enquote {\bibinfo {title} {Necessary and
  sufficient quantum information characterization of einstein-podolsky-rosen
  steering},}\ }\href {\doibase 10.1103/PhysRevLett.114.060404} {\bibfield
  {journal} {\bibinfo  {journal} {Phys. Rev. Lett.}\ }\textbf {\bibinfo
  {volume} {114}},\ \bibinfo {pages} {060404} (\bibinfo {year}
  {2015})}\BibitemShut {NoStop}%
\bibitem [{\citenamefont {Cavalcanti}\ \emph {et~al.}(2015)\citenamefont
  {Cavalcanti}, \citenamefont {Broadbent}, \citenamefont {Walborn},\ and\
  \citenamefont {Wiseman}}]{CBW+15}%
  \BibitemOpen
  \bibfield  {author} {\bibinfo {author} {\bibfnamefont {Eric~G.}\ \bibnamefont
  {Cavalcanti}}, \bibinfo {author} {\bibfnamefont {Curtis~J.}\ \bibnamefont
  {Broadbent}}, \bibinfo {author} {\bibfnamefont {Stephen~P.}\ \bibnamefont
  {Walborn}}, \ and\ \bibinfo {author} {\bibfnamefont {Howard~M.}\ \bibnamefont
  {Wiseman}},\ }\bibfield  {title} {\enquote {\bibinfo {title} {80 years of
  steering and the einstein-podolsky-rosen paradox: introduction},}\ }\href
  {\doibase 10.1364/JOSAB.32.00EPR1} {\bibfield  {journal} {\bibinfo  {journal}
  {J. Opt. Soc. Am. B}\ }\textbf {\bibinfo {volume} {32}},\ \bibinfo {pages}
  {EPR1--EPR2} (\bibinfo {year} {2015})}\BibitemShut {NoStop}%
\bibitem [{\citenamefont {Gallego}\ and\ \citenamefont {Aolita}(2015)}]{GA15}%
  \BibitemOpen
  \bibfield  {author} {\bibinfo {author} {\bibfnamefont {Rodrigo}\ \bibnamefont
  {Gallego}}\ and\ \bibinfo {author} {\bibfnamefont {Leandro}\ \bibnamefont
  {Aolita}},\ }\bibfield  {title} {\enquote {\bibinfo {title} {Resource theory
  of steering},}\ }\href {\doibase 10.1103/PhysRevX.5.041008} {\bibfield
  {journal} {\bibinfo  {journal} {Phys. Rev. X}\ }\textbf {\bibinfo {volume}
  {5}},\ \bibinfo {pages} {041008} (\bibinfo {year} {2015})}\BibitemShut
  {NoStop}%
\bibitem [{\citenamefont {Zhu}\ \emph {et~al.}(2016)\citenamefont {Zhu},
  \citenamefont {Hayashi},\ and\ \citenamefont {Chen}}]{ZHC16}%
  \BibitemOpen
  \bibfield  {author} {\bibinfo {author} {\bibfnamefont {Huangjun}\
  \bibnamefont {Zhu}}, \bibinfo {author} {\bibfnamefont {Masahito}\
  \bibnamefont {Hayashi}}, \ and\ \bibinfo {author} {\bibfnamefont {Lin}\
  \bibnamefont {Chen}},\ }\bibfield  {title} {\enquote {\bibinfo {title}
  {Universal steering criteria},}\ }\href {\doibase
  10.1103/PhysRevLett.116.070403} {\bibfield  {journal} {\bibinfo  {journal}
  {Phys. Rev. Lett.}\ }\textbf {\bibinfo {volume} {116}},\ \bibinfo {pages}
  {070403} (\bibinfo {year} {2016})}\BibitemShut {NoStop}%
\bibitem [{\citenamefont {Law}\ \emph {et~al.}(2014)\citenamefont {Law},
  \citenamefont {Thinh}, \citenamefont {Bancal},\ and\ \citenamefont
  {Scarani}}]{LTB+14}%
  \BibitemOpen
  \bibfield  {author} {\bibinfo {author} {\bibfnamefont {Yun~Zhi}\ \bibnamefont
  {Law}}, \bibinfo {author} {\bibfnamefont {Le~Phuc}\ \bibnamefont {Thinh}},
  \bibinfo {author} {\bibfnamefont {Jean-Daniel}\ \bibnamefont {Bancal}}, \
  and\ \bibinfo {author} {\bibfnamefont {Valerio}\ \bibnamefont {Scarani}},\
  }\bibfield  {title} {\enquote {\bibinfo {title} {Quantum randomness
  extraction for various levels of characterization of the devices},}\ }\href
  {http://stacks.iop.org/1751-8121/47/i=42/a=424028} {\bibfield  {journal}
  {\bibinfo  {journal} {Journal of Physics A: Mathematical and Theoretical}\
  }\textbf {\bibinfo {volume} {47}},\ \bibinfo {pages} {424028} (\bibinfo
  {year} {2014})}\BibitemShut {NoStop}%
\bibitem [{\citenamefont {Hsieh}\ \emph {et~al.}(2016)\citenamefont {Hsieh},
  \citenamefont {Liang},\ and\ \citenamefont {Lee}}]{HLL16}%
  \BibitemOpen
  \bibfield  {author} {\bibinfo {author} {\bibfnamefont {Chung-Yun}\
  \bibnamefont {Hsieh}}, \bibinfo {author} {\bibfnamefont {Yeong-Cherng}\
  \bibnamefont {Liang}}, \ and\ \bibinfo {author} {\bibfnamefont {Ray-Kuang}\
  \bibnamefont {Lee}},\ }\bibfield  {title} {\enquote {\bibinfo {title}
  {Quantum steerability: Characterization, quantification, superactivation, and
  unbounded amplification},}\ }\href {\doibase 10.1103/PhysRevA.94.062120}
  {\bibfield  {journal} {\bibinfo  {journal} {Phys. Rev. A}\ }\textbf {\bibinfo
  {volume} {94}},\ \bibinfo {pages} {062120} (\bibinfo {year}
  {2016})}\BibitemShut {NoStop}%
\bibitem [{\citenamefont {Kaur}\ \emph {et~al.}(2017)\citenamefont {Kaur},
  \citenamefont {Wang},\ and\ \citenamefont {Wilde}}]{KWW17}%
  \BibitemOpen
  \bibfield  {author} {\bibinfo {author} {\bibfnamefont {Eneet}\ \bibnamefont
  {Kaur}}, \bibinfo {author} {\bibfnamefont {Xiaoting}\ \bibnamefont {Wang}}, \
  and\ \bibinfo {author} {\bibfnamefont {Mark~M.}\ \bibnamefont {Wilde}},\
  }\bibfield  {title} {\enquote {\bibinfo {title} {Conditional mutual
  information and quantum steering},}\ }\href {\doibase
  10.1103/PhysRevA.96.022332} {\bibfield  {journal} {\bibinfo  {journal} {Phys.
  Rev. A}\ }\textbf {\bibinfo {volume} {96}},\ \bibinfo {pages} {022332}
  (\bibinfo {year} {2017})}\BibitemShut {NoStop}%
\bibitem [{\citenamefont {Kaur}\ and\ \citenamefont {Wilde}(2017)}]{KW17}%
  \BibitemOpen
  \bibfield  {author} {\bibinfo {author} {\bibfnamefont {Eneet}\ \bibnamefont
  {Kaur}}\ and\ \bibinfo {author} {\bibfnamefont {Mark~M}\ \bibnamefont
  {Wilde}},\ }\bibfield  {title} {\enquote {\bibinfo {title} {Relative entropy
  of steering: on its definition and properties},}\ }\href
  {http://stacks.iop.org/1751-8121/50/i=46/a=465301} {\bibfield  {journal}
  {\bibinfo  {journal} {Journal of Physics A: Mathematical and Theoretical}\
  }\textbf {\bibinfo {volume} {50}},\ \bibinfo {pages} {465301} (\bibinfo
  {year} {2017})}\BibitemShut {NoStop}%
\bibitem [{\citenamefont {Das}\ \emph {et~al.}(2018)\citenamefont {Das},
  \citenamefont {Datta}, \citenamefont {Jebaratnam},\ and\ \citenamefont
  {Majumdar}}]{DDJ+18}%
  \BibitemOpen
  \bibfield  {author} {\bibinfo {author} {\bibfnamefont {Debarshi}\
  \bibnamefont {Das}}, \bibinfo {author} {\bibfnamefont {Shounak}\ \bibnamefont
  {Datta}}, \bibinfo {author} {\bibfnamefont {C.}~\bibnamefont {Jebaratnam}}, \
  and\ \bibinfo {author} {\bibfnamefont {A.~S.}\ \bibnamefont {Majumdar}},\
  }\bibfield  {title} {\enquote {\bibinfo {title} {Cost of
  einstein-podolsky-rosen steering in the context of extremal boxes},}\ }\href
  {\doibase 10.1103/PhysRevA.97.022110} {\bibfield  {journal} {\bibinfo
  {journal} {Phys. Rev. A}\ }\textbf {\bibinfo {volume} {97}},\ \bibinfo
  {pages} {022110} (\bibinfo {year} {2018})}\BibitemShut {NoStop}%
\bibitem [{\citenamefont {Schneeloch}\ and\ \citenamefont
  {Howland}(2018)}]{SH18}%
  \BibitemOpen
  \bibfield  {author} {\bibinfo {author} {\bibfnamefont {James}\ \bibnamefont
  {Schneeloch}}\ and\ \bibinfo {author} {\bibfnamefont {Gregory~A.}\
  \bibnamefont {Howland}},\ }\bibfield  {title} {\enquote {\bibinfo {title}
  {Quantifying high-dimensional entanglement with einstein-podolsky-rosen
  correlations},}\ }\href {\doibase 10.1103/PhysRevA.97.042338} {\bibfield
  {journal} {\bibinfo  {journal} {Phys. Rev. A}\ }\textbf {\bibinfo {volume}
  {97}},\ \bibinfo {pages} {042338} (\bibinfo {year} {2018})}\BibitemShut
  {NoStop}%
\bibitem [{\citenamefont {Quintino}\ \emph {et~al.}(2015)\citenamefont
  {Quintino}, \citenamefont {V\'ertesi}, \citenamefont {Cavalcanti},
  \citenamefont {Augusiak}, \citenamefont {Demianowicz}, \citenamefont
  {Ac\'{\i}n},\ and\ \citenamefont {Brunner}}]{QVC+15}%
  \BibitemOpen
  \bibfield  {author} {\bibinfo {author} {\bibfnamefont {Marco~T\'ulio}\
  \bibnamefont {Quintino}}, \bibinfo {author} {\bibfnamefont {Tam\'as}\
  \bibnamefont {V\'ertesi}}, \bibinfo {author} {\bibfnamefont {Daniel}\
  \bibnamefont {Cavalcanti}}, \bibinfo {author} {\bibfnamefont {Remigiusz}\
  \bibnamefont {Augusiak}}, \bibinfo {author} {\bibfnamefont {Maciej}\
  \bibnamefont {Demianowicz}}, \bibinfo {author} {\bibfnamefont {Antonio}\
  \bibnamefont {Ac\'{\i}n}}, \ and\ \bibinfo {author} {\bibfnamefont {Nicolas}\
  \bibnamefont {Brunner}},\ }\bibfield  {title} {\enquote {\bibinfo {title}
  {Inequivalence of entanglement, steering, and bell nonlocality for general
  measurements},}\ }\href {\doibase 10.1103/PhysRevA.92.032107} {\bibfield
  {journal} {\bibinfo  {journal} {Phys. Rev. A}\ }\textbf {\bibinfo {volume}
  {92}},\ \bibinfo {pages} {032107} (\bibinfo {year} {2015})}\BibitemShut
  {NoStop}%
\bibitem [{\citenamefont {Costa}\ and\ \citenamefont {Angelo}(2016)}]{CA16}%
  \BibitemOpen
  \bibfield  {author} {\bibinfo {author} {\bibfnamefont {A.~C.~S.}\
  \bibnamefont {Costa}}\ and\ \bibinfo {author} {\bibfnamefont {R.~M.}\
  \bibnamefont {Angelo}},\ }\bibfield  {title} {\enquote {\bibinfo {title}
  {Quantification of einstein-podolski-rosen steering for two-qubit states},}\
  }\href {\doibase 10.1103/PhysRevA.93.020103} {\bibfield  {journal} {\bibinfo
  {journal} {Phys. Rev. A}\ }\textbf {\bibinfo {volume} {93}},\ \bibinfo
  {pages} {020103} (\bibinfo {year} {2016})}\BibitemShut {NoStop}%
\bibitem [{\citenamefont {Guo}\ and\ \citenamefont {Wu}(2014)}]{GW14}%
  \BibitemOpen
  \bibfield  {author} {\bibinfo {author} {\bibfnamefont {Y}~\bibnamefont
  {Guo}}\ and\ \bibinfo {author} {\bibfnamefont {S}~\bibnamefont {Wu}},\
  }\bibfield  {title} {\enquote {\bibinfo {title} {Quantum correlation exists
  in any non-product state},}\ }\href {\doibase 10.1038/srep07179} {\bibfield
  {journal} {\bibinfo  {journal} {Sci. Rep}\ }\textbf {\bibinfo {volume} {4}},\
  \bibinfo {pages} {7179} (\bibinfo {year} {2014})}\BibitemShut {NoStop}%
\bibitem [{\citenamefont {Nielsen}\ and\ \citenamefont {Chuang}(2000)}]{NC00}%
  \BibitemOpen
  \bibfield  {author} {\bibinfo {author} {\bibfnamefont {M.~A.}\ \bibnamefont
  {Nielsen}}\ and\ \bibinfo {author} {\bibfnamefont {I.~L.}\ \bibnamefont
  {Chuang}},\ }\href@noop {} {\emph {\bibinfo {title} {Quantum Computation and
  Quantum Information}}}\ (\bibinfo  {publisher} {Cambridge University Press,
  Cambridge, England},\ \bibinfo {year} {2000})\BibitemShut {NoStop}%
\bibitem [{\citenamefont {Wu}\ \emph {et~al.}(2014)\citenamefont {Wu},
  \citenamefont {Ma}, \citenamefont {Chen},\ and\ \citenamefont {Yu}}]{WMC+14}%
  \BibitemOpen
  \bibfield  {author} {\bibinfo {author} {\bibfnamefont {S}~\bibnamefont {Wu}},
  \bibinfo {author} {\bibfnamefont {Z}~\bibnamefont {Ma}}, \bibinfo {author}
  {\bibfnamefont {Z}~\bibnamefont {Chen}}, \ and\ \bibinfo {author}
  {\bibfnamefont {S}~\bibnamefont {Yu}},\ }\bibfield  {title} {\enquote
  {\bibinfo {title} {Reveal quantum correlation in complementary bases},}\
  }\href {\doibase doi:10.1038/srep04036} {\bibfield  {journal} {\bibinfo
  {journal} {Sci. Rep}\ }\textbf {\bibinfo {volume} {4}},\ \bibinfo {pages}
  {4036} (\bibinfo {year} {2014})}\BibitemShut {NoStop}%
\bibitem [{\citenamefont {Bell}(1964)}]{Bel64}%
  \BibitemOpen
  \bibfield  {author} {\bibinfo {author} {\bibfnamefont {J.~S.}\ \bibnamefont
  {Bell}},\ }\bibfield  {title} {\enquote {\bibinfo {title} {On the einstein
  podolsky rosen paradox},}\ }\href {\doibase
  10.1103/PhysicsPhysiqueFizika.1.195} {\bibfield  {journal} {\bibinfo
  {journal} {Physics Physique Fizika}\ }\textbf {\bibinfo {volume} {1}},\
  \bibinfo {pages} {195--200} (\bibinfo {year} {1964})}\BibitemShut {NoStop}%
\bibitem [{\citenamefont {Brunner}\ \emph {et~al.}(2014)\citenamefont
  {Brunner}, \citenamefont {Cavalcanti}, \citenamefont {Pironio}, \citenamefont
  {Scarani},\ and\ \citenamefont {Wehner}}]{BCP+14}%
  \BibitemOpen
  \bibfield  {author} {\bibinfo {author} {\bibfnamefont {Nicolas}\ \bibnamefont
  {Brunner}}, \bibinfo {author} {\bibfnamefont {Daniel}\ \bibnamefont
  {Cavalcanti}}, \bibinfo {author} {\bibfnamefont {Stefano}\ \bibnamefont
  {Pironio}}, \bibinfo {author} {\bibfnamefont {Valerio}\ \bibnamefont
  {Scarani}}, \ and\ \bibinfo {author} {\bibfnamefont {Stephanie}\ \bibnamefont
  {Wehner}},\ }\bibfield  {title} {\enquote {\bibinfo {title} {Bell
  nonlocality},}\ }\href {\doibase 10.1103/RevModPhys.86.419} {\bibfield
  {journal} {\bibinfo  {journal} {Rev. Mod. Phys.}\ }\textbf {\bibinfo {volume}
  {86}},\ \bibinfo {pages} {419--478} (\bibinfo {year} {2014})}\BibitemShut
  {NoStop}%
\bibitem [{\citenamefont {Cavalcanti}\ \emph {et~al.}(2009)\citenamefont
  {Cavalcanti}, \citenamefont {Jones}, \citenamefont {Wiseman},\ and\
  \citenamefont {Reid}}]{CJW+09}%
  \BibitemOpen
  \bibfield  {author} {\bibinfo {author} {\bibfnamefont {E.~G.}\ \bibnamefont
  {Cavalcanti}}, \bibinfo {author} {\bibfnamefont {S.~J.}\ \bibnamefont
  {Jones}}, \bibinfo {author} {\bibfnamefont {H.~M.}\ \bibnamefont {Wiseman}},
  \ and\ \bibinfo {author} {\bibfnamefont {M.~D.}\ \bibnamefont {Reid}},\
  }\bibfield  {title} {\enquote {\bibinfo {title} {Experimental criteria for
  steering and the einstein-podolsky-rosen paradox},}\ }\href {\doibase
  10.1103/PhysRevA.80.032112} {\bibfield  {journal} {\bibinfo  {journal} {Phys.
  Rev. A}\ }\textbf {\bibinfo {volume} {80}},\ \bibinfo {pages} {032112}
  (\bibinfo {year} {2009})}\BibitemShut {NoStop}%
\bibitem [{\citenamefont {Luo}(2008)}]{Luo08}%
  \BibitemOpen
  \bibfield  {author} {\bibinfo {author} {\bibfnamefont {Shunlong}\
  \bibnamefont {Luo}},\ }\bibfield  {title} {\enquote {\bibinfo {title}
  {Quantum discord for two-qubit systems},}\ }\href {\doibase
  10.1103/PhysRevA.77.042303} {\bibfield  {journal} {\bibinfo  {journal} {Phys.
  Rev. A}\ }\textbf {\bibinfo {volume} {77}},\ \bibinfo {pages} {042303}
  (\bibinfo {year} {2008})}\BibitemShut {NoStop}%
\bibitem [{\citenamefont {Henderson}\ and\ \citenamefont
  {Vedral}(2001)}]{HV01}%
  \BibitemOpen
  \bibfield  {author} {\bibinfo {author} {\bibfnamefont {L}~\bibnamefont
  {Henderson}}\ and\ \bibinfo {author} {\bibfnamefont {V}~\bibnamefont
  {Vedral}},\ }\bibfield  {title} {\enquote {\bibinfo {title} {Classical,
  quantum and total correlations},}\ }\href
  {http://stacks.iop.org/0305-4470/34/i=35/a=315} {\bibfield  {journal}
  {\bibinfo  {journal} {Journal of Physics A: Mathematical and General}\
  }\textbf {\bibinfo {volume} {34}},\ \bibinfo {pages} {6899} (\bibinfo {year}
  {2001})}\BibitemShut {NoStop}%
\bibitem [{\citenamefont {Uola}\ \emph {et~al.}(2015)\citenamefont {Uola},
  \citenamefont {Budroni}, \citenamefont {G\"uhne},\ and\ \citenamefont
  {Pellonp\"a\"a}}]{UBG+15}%
  \BibitemOpen
  \bibfield  {author} {\bibinfo {author} {\bibfnamefont {Roope}\ \bibnamefont
  {Uola}}, \bibinfo {author} {\bibfnamefont {Costantino}\ \bibnamefont
  {Budroni}}, \bibinfo {author} {\bibfnamefont {Otfried}\ \bibnamefont
  {G\"uhne}}, \ and\ \bibinfo {author} {\bibfnamefont {Juha-Pekka}\
  \bibnamefont {Pellonp\"a\"a}},\ }\bibfield  {title} {\enquote {\bibinfo
  {title} {One-to-one mapping between steering and joint measurability
  problems},}\ }\href {\doibase 10.1103/PhysRevLett.115.230402} {\bibfield
  {journal} {\bibinfo  {journal} {Phys. Rev. Lett.}\ }\textbf {\bibinfo
  {volume} {115}},\ \bibinfo {pages} {230402} (\bibinfo {year}
  {2015})}\BibitemShut {NoStop}%
\bibitem [{\citenamefont {Chen}\ \emph {et~al.}(2016)\citenamefont {Chen},
  \citenamefont {Budroni}, \citenamefont {Liang},\ and\ \citenamefont
  {Chen}}]{CBL+16}%
  \BibitemOpen
  \bibfield  {author} {\bibinfo {author} {\bibfnamefont {Shin-Liang}\
  \bibnamefont {Chen}}, \bibinfo {author} {\bibfnamefont {Costantino}\
  \bibnamefont {Budroni}}, \bibinfo {author} {\bibfnamefont {Yeong-Cherng}\
  \bibnamefont {Liang}}, \ and\ \bibinfo {author} {\bibfnamefont {Yueh-Nan}\
  \bibnamefont {Chen}},\ }\bibfield  {title} {\enquote {\bibinfo {title}
  {Natural framework for device-independent quantification of quantum
  steerability, measurement incompatibility, and self-testing},}\ }\href
  {\doibase 10.1103/PhysRevLett.116.240401} {\bibfield  {journal} {\bibinfo
  {journal} {Phys. Rev. Lett.}\ }\textbf {\bibinfo {volume} {116}},\ \bibinfo
  {pages} {240401} (\bibinfo {year} {2016})}\BibitemShut {NoStop}%
\end{thebibliography}%
\end{document}